\RequirePackage{fix-cm}
\documentclass[smallextended]{svjour3}       
\smartqed  
\usepackage{graphicx}
\usepackage{color,pstcol}
\newcommand{\Ysssrcnb}
           {M}

\newcommand{\Yssbothtimevalnb}
           {N}

\newcommand{\Ysswayindexfourth}
           {\ell}

\newcommand{\Ysssrcsigdisctimecentmatrix}
           {S}
\newcommand{\ysssrcsigdisctimecentvec}
          {$\Ysssrcsigdisctimecentvec$}
\newcommand{\Ysssrcsigdisctimecentvec}
           {s}

\newcommand{\ysssrcsigdisctimecentone}
          {$\Ysssrcsigdisctimecentone$}
\newcommand{\Ysssrcsigdisctimecentone}
           {\Ysssrcsigdisctimecentvec _{1}}

\newcommand{\Ysssrcsigdisctimecenttwo}
           {\Ysssrcsigdisctimecentvec _{2}}

\newcommand{\Ysssrcsigdisctimecentthree}
           {\Ysssrcsigdisctimecentvec _{3}}

\newcommand{\ysssrcsigdisctimecentlast}
          {$\Ysssrcsigdisctimecentlast$}
\newcommand{\Ysssrcsigdisctimecentlast}
           {\Ysssrcsigdisctimecentvec _{\Ysssrcnb}}

\newcommand{\Yssextquadsrcsigdisctimecentmatrix}
           {
\widetilde
{\Ysssrcsigdisctimecentmatrix}}

\newcommand{\Yssextquadsrcsigdisctimecentmatrixestim}
           {\widehat{\Yssextquadsrcsigdisctimecentmatrix}}
\newcommand{\Yssextquadsrcsigdisctimecentmatrixestimnotvecallsampsrcnot}
{\check{s}}
\newcommand{\yssmixsigdisctimecentvec}
          {$\Yssmixsigdisctimecentvec$}
\newcommand{\Yssmixsigdisctimecentvec}
           {x}

\newcommand{\yssmixsigdisctimecentone}
          {$\Yssmixsigdisctimecentone$}
\newcommand{\Yssmixsigdisctimecentone}
           {\Yssmixsigdisctimecentvec _{1}}

\newcommand{\Yssmixsigdisctimecenttwo}
           {\Yssmixsigdisctimecentvec _{2}}

\newcommand{\Yssmixsigdisctimecentthree}
{\Yssmixsigdisctimecentvec _{3}}

\newcommand{\yssoutsepsystsigdisctimecentvec}
          {$\Yssoutsepsystsigdisctimecentvec$}
\newcommand{\Yssoutsepsystsigdisctimecentvec}
           {y}

\newcommand{\yssoutsepsystsigdisctimecentone}
          {$\Yssoutsepsystsigdisctimecentone$}
\newcommand{\Yssoutsepsystsigdisctimecentone}
           {\Yssoutsepsystsigdisctimecentvec _{1}}

\newcommand{\Yssmixmatrixscalar}
           {A}

\newcommand{\Yssmixmatrixscalarquad}
{B}

\newcommand{\Yssmixmatrixscalarelquadnoindex}
          {b}

\newcommand{\Yssmixmatrixscalarelquadnoindexestim}
          {\hat{\Yssmixmatrixscalarelquadnoindex}}

\newcommand{\Yssmixmatrixscalarquadext}
           {
\widetilde
{\Yssmixmatrixscalar}}
\newcommand{\Yssmixmatrixscalarquadextestim}
           {\widehat{\Yssmixmatrixscalarquadext}}

\newcommand{\Ysssepsystmatrixscalar}
{C}

\newcommand{\Ysssepsystmatrixscalarelnoindex}
           {c}

\newcommand{\Yqubitonesevspace}
{{\cal E}}
\newcommand{\yqubitoneindexstd}
{$\Yqubitoneindexstd$}
\newcommand{\Yqubitoneindexstd}
{j}
\newcommand{\yqubitonetimeinit}
{$\Yqubitonetimeinit$}
\newcommand{\Yqubitonetimeinit}
{t_0}
\newcommand{\yqubitonetimefinal}
{$\Yqubitonetimefinal$}
\newcommand{\Yqubitonetimefinal}
{t}
\newcommand{\yqubitonetimeinitstateindexone}
{$\Yqubitonetimeinitstateindexone$}
\newcommand{\Yqubitonetimeinitstateindexone}
{| \psi_1 ( \Yqubitonetimeinit ) \rangle}
\newcommand{\yqubitonetimeinitstateindextwo}
{$\Yqubitonetimeinitstateindextwo$}
\newcommand{\Yqubitonetimeinitstateindextwo}
{| \psi_2 ( \Yqubitonetimeinit ) \rangle}
\newcommand{\ymixsyststateinitial}
{$\Ymixsyststateinitial$}
\newcommand{\Ymixsyststateinitial}
{| \psi 
( \Yqubitonetimeinit ) \rangle
}
\newcommand{\Ytwoqubitwritereadtimeinterval}
{\tau}

\newcommand{\Yqubitbothtimeinitstatecoefindexequalstd}
{k}

\newcommand{\Yqubitbothtimefinalstatecoefindexindexstd}
{k}
\newcommand{\ymixsyststatefinal}
{$\Ymixsyststatefinal$}
\newcommand{\Ymixsyststatefinal}
{| \psi (t) 
\rangle
}
\newcommand{\Yopmixbases}
{Q}
\newcommand{\Yopmixbasesimagin}
{\Yopmixbases_{\Ysqrtminusone}}

\newcommand{\Ysepsystdedicstateoutcoefnot}
{c}
\newcommand{\Ysepsystdedicstateoutcoefpart}
{\Ysepsystdedicstateoutcoefnot_{5}}

\newcommand{\Ysepsystdedicstateoutcoefprobnot}
{P}

\newcommand{\Ysepsystdedicstateoutcoefprobnotox}
{P}

\newcommand{\Ysepsystdedicopmixbasesonestateoutcoefprobzznot}
{\Ysepsystdedicstateoutcoefprobnot}

\newcommand{\Ysepsystdedicopmixbasesimaginonestateoutcoefprobzznot}
{\Ysepsystdedicstateoutcoefprobnot}
\newcommand{\ysepsystdedicopmixbasesimaginonestateoutcoefprobzzplusplus}
{$\Ysepsystdedicopmixbasesimaginonestateoutcoefprobzzplusplus$}
\newcommand{\Ysepsystdedicopmixbasesimaginonestateoutcoefprobzzplusplus}
{\Ysepsystdedicopmixbasesimaginonestateoutcoefprobzznot
_{1z}
( \Yopmixbasesimagin )
}

\newcommand{\Yindexstdforsepmixdiagel}
{k}

\newcommand{\Yqubitindexstd}
{j}

\newcommand{\Yqubitnbarbstatecoefindexequalstd}
{k}
\newcommand{\ytwoqubitsprobaplusplus}
{$\Ytwoqubitsprobaplusplus$}
\newcommand{\Ytwoqubitsprobaplusplus}
{p_1}
\newcommand{\ytwoqubitsprobaplusminus}
{$\Ytwoqubitsprobaplusminus$}
\newcommand{\Ytwoqubitsprobaplusminus}
{p_2}
\newcommand{\ytwoqubitsprobaminusplus}
{$\Ytwoqubitsprobaminusplus$}
\newcommand{\Ytwoqubitsprobaminusplus}
{p_3}
\newcommand{\ytwoqubitsprobaminusminus}
{$\Ytwoqubitsprobaminusminus$}
\newcommand{\Ytwoqubitsprobaminusminus}
{p_4}
\newcommand{\ytwoqubitsprobaindexstd}
{$\Ytwoqubitsprobaindexstd$}
\newcommand{\Ytwoqubitsprobaindexstd}
{\Ytwoqubitsprobanot
_{\Yqubitbothtimefinalstatecoefindexindexstd}
}
\newcommand{\Ytwoqubitsprobanot}
{p}
\newcommand{\ytwoqubitresultphaseinit}
{$\Ytwoqubitresultphaseinit$}
\newcommand{\Ytwoqubitresultphaseinit}
{\Delta _I}
\newcommand{\Ytwoqubitresultphaseevol}
{\Delta _E}
\newcommand{\ytwoqubitresultphaseevolsin}
{$\Ytwoqubitresultphaseevolsin$}
\newcommand{\Ytwoqubitresultphaseevolsin}
{v}

\newcommand{\Ytwoqubitresultphaseevolsinestim}
{\overline{\Ytwoqubitresultphaseevolsin}}

\newcommand{\Ytwoqubitresultphaseevolsinestimtwo}
{\widehat{\Ytwoqubitresultphaseevolsin}}

\newcommand{\Yparamqubitbothstateplusmodulusnot}
{r}
\newcommand{\Yparamqubitonestateplusmodulus}
{{\Yparamqubitbothstateplusmodulusnot}_1}
\newcommand{\Yparamqubittwostateplusmodulus}
{{\Yparamqubitbothstateplusmodulusnot}_2}
\newcommand{\yparamqubitindexstdstateplusmodulus}
{$\Yparamqubitindexstdstateplusmodulus$}
\newcommand{\Yparamqubitindexstdstateplusmodulus}
{{\Yparamqubitbothstateplusmodulusnot}_{\Yqubitindexstd}}
\newcommand{\Yparamqubitbothstateminusmodulusnot}
{q}
\newcommand{\Yparamqubitonestateminusmodulus}
{{\Yparamqubitbothstateminusmodulusnot}_1}
\newcommand{\Yparamqubittwostateminusmodulus}
{{\Yparamqubitbothstateminusmodulusnot}_2}
\newcommand{\yparamqubitindexstdstateminusmodulus}
{$\Yparamqubitindexstdstateminusmodulus$}
\newcommand{\Yparamqubitindexstdstateminusmodulus}
{{\Yparamqubitbothstateminusmodulusnot}_{\Yqubitindexstd}}
\newcommand{\Yparamqubitbothstateplusphasenot}
{\theta}
\newcommand{\Yparamqubitonestateplusphase}
{\Yparamqubitbothstateplusphasenot_1}
\newcommand{\Yparamqubittwostateplusphase}
{\Yparamqubitbothstateplusphasenot_2}
\newcommand{\yparamqubitindexstdstateplusphase}
{$\Yparamqubitindexstdstateplusphase$}
\newcommand{\Yparamqubitindexstdstateplusphase}
{{\Yparamqubitbothstateplusphasenot}_{\Yqubitindexstd}}
\newcommand{\Yparamqubitbothstateminusphasenot}
{\phi}
\newcommand{\Yparamqubitonestateminusphase}
{\Yparamqubitbothstateminusphasenot_1}
\newcommand{\Yparamqubittwostateminusphase}
{\Yparamqubitbothstateminusphasenot_2}
\newcommand{\yparamqubitindexstdstateminusphase}
{$\Yparamqubitindexstdstateminusphase$}
\newcommand{\Yparamqubitindexstdstateminusphase}
{{\Yparamqubitbothstateminusphasenot}_{\Yqubitindexstd}}
\newcommand{\Ymagfieldnot}
{B}
\newcommand{\Ymagfieldvec}
{\overrightarrow{\Ymagfieldnot}}
\newcommand{\Yhamiltonfieldscale}
{G}
\newcommand{\yexchangetensorppalvaluexy}
{$\Yexchangetensorppalvaluexy$}
\newcommand{\Yexchangetensorppalvaluexy}
{J_{xy}}

\newcommand{\ysqrtminusone}
{$\Ysqrtminusone$}
\newcommand{\Ysqrtminusone}
{i}

\newcommand{\Yprob}{P}

\newcommand{\Ymixmodelstateoutqubitindexonedirectionnonerv}
{
b
_{
1
}
}

\newcommand{\Ymixmodelstateoutqubitindextwodirectionnonerv}
{
b
_{
2
}
}
\newcommand{\ymixmodelstateoutqubitindexstddirectionnonerv}
{$\Ymixmodelstateoutqubitindexstddirectionnonerv$}
\newcommand{\Ymixmodelstateoutqubitindexstddirectionnonerv}
{
b
_{
\Yqubitindexstd
}
}
\newcommand{\ytextmodifartitihundredsixtyfourversiononestepone}[1]
{#1}
\newcommand{\ytextmodifartitihundredsixtyfourversiononesteptwo}[1]
{#1}
\newcommand{\ytextmodifartitihundredsixtyfourversiononestepthree}[1]
{#1}
\journalname{}
\begin{document}

\title{%
Statistical
intrusion detection and
eavesdropping
in 
quantum channels with 
coupling:
Multiple-preparation 
and single-preparation methods%
}

\titlerunning{Intrusion detection and eavesdroping methods
for quantum channels}

\author{Yannick Deville
\and
Alain Deville
\and
Ali Mansour
\and
Razvan Scripcaru
\and
Cornel Ioana}

\institute{%
Y. Deville
\at
Universit\'e de Toulouse,
UPS, CNRS, CNES, OMP,
IRAP,
Toulouse, France\\
Tel.: +33561332824\\
Fax: +33561332840\\
\email{yannick.deville@irap.omp.eu}\\
ORCID:
0000-0002-8769-2446
\and
A. Deville
\at
Aix-Marseille Universit\'e, CNRS,
IM2NP UMR 7334,
Marseille, France\\
ORCID: 0000-0001-5246-8391
\and
A.
Mansour
\at
ENSTA Bretagne, 
Lab-STICC, UMR 6285,
Brest, France\\
ORCID: 0000-0003-4144-8832
\and
R. Scripcaru
\at
Universite 
Grenoble Alpes, 
GIPSA-lab, UMR 5216, 
Saint 
Martin d'Heres, France
\and
C. Ioana
\at
Universite 
Grenoble Alpes, 
GIPSA-lab, UMR 5216, 
Saint 
Martin d'Heres, France\\
ORCID: 0000-0001-6581-3000}

\date{Received: date / Accepted: date}

\maketitle

\begin{abstract}
Classical, i.e. non-quantum, communications include 
configurations with
multiple-input
multiple-output (MIMO) channels.
Some associated signal processing tasks consider
these
channels in a symmetric way, i.e. by assigning
the same
role to all channel inputs, and 
\ytextmodifartitihundredsixtyfourversiononestepthree{similarly
to all channel
outputs.}
These tasks especially include channel 
\ytextmodifartitihundredsixtyfourversiononestepthree{identification/estimation}
and channel equalization, 
tightly
connected 
with
source separation.
Their most challenging version is the blind one,
i.e. when the receivers have (almost) no prior
knowledge about the emitted signals.
Other signal processing tasks 
consider
classical communication
channels in an asymmetric way.
This especially includes the situation when data
are sent by Emitter 1 to Receiver 1 through a main
channel, and an 
``intruder'' 
(including
Receiver 2)
interferes with that
channel so as to extract information, thus
performing so-called eavesdropping,
while
Receiver 1 may aim at detecting that intrusion.
Part of the above processing tasks have been
extended to quantum channels, 
including those that
have several
quantum bits (qubits) at their
input and output.
\ytextmodifartitihundredsixtyfourversiononestepthree{For 
such quantum channels,
beyond previously reported work for symmetric scenarios,
we here address asymmetric (blind and non-blind) ones,}
with emphasis
on intrusion detection and additional comments
about eavesdropping.
\ytextmodifartitihundredsixtyfourversiononestepthree{To develop 
fundamental concepts,
we first consider channels with exchange coupling as a toy model.
We 
especially use
the general
quantum information processing
framework
that we recently developed,
to derive new attractive intrusion detection methods
based on a single preparation of each state.}
Finally, we
discuss
how 
the proposed
methods
might be extended,
beyond the specific class of channels
analyzed here.
\end{abstract}

\section{%
Previous works
and
problem statement}
\label{sec-intro}
The communication systems 
that involve classical, i.e. non-quantum,
channels 
give rise to a variety of signal processing
problems. Among them, two closely related problems
are (i) channel identification, i.e. estimation, and
(ii) channel equalization
\cite{livreproakis,book-equalization-ding-li}.
These problems may
be seen as the communication
version of, respectively,
(i) system identification
and
(ii)
system inversion and its multiple-signal
extension to source separation,
that are generic signal processing problems,
which include various versions defined as follows.
The
basic 
version of system identification 
addresses
single-input single-output (SISO)
systems.
It
consists in
estimating the unknown parameter values of such a 
system
{(i.e. of its transform)}
belonging to a known class, by using known values of
its input 
(source signal
\ysssrcsigdisctimecentvec )
and output
(%
signal 
\yssmixsigdisctimecentvec ).
This version
{is}
stated to be non-blind
(or supervised),
as opposed to the more challenging, blind
(or unsupervised),
version of
that problem, where the input values are unknown (%
{and uncontrolled%
,%
}
but the
input signal may be known to belong to a given
class)%
{: 
See%
}
\cite{a593}.
Both versions may then be extended 
to
multiple-input multiple-output (MIMO) systems.

Besides, in various applications, what is needed is not
the direct transform achieved by the above system, but the inverse
of that transform (assuming it is invertible).
For SISO non-blind and  blind configurations,
this is motivated by the fact that one eventually
only accesses the ouput
\yssmixsigdisctimecentvec\
of the above direct system, and one aims at deriving a signal
\yssoutsepsystsigdisctimecentvec\
which ideally restores the original
source signal
\ysssrcsigdisctimecentvec .
To this end, one may 
use the above-mentioned 
system identification methods in order to first estimate the direct system,
then derive its inverse
and eventually transfer the 
output 
\yssmixsigdisctimecentvec\
of the direct system through the inverse system.
Alternatively, one may develop
methods for initially
identifying the \emph{inverse} system itself.
Extended versions 
of
this ``%
(unknown)
system inversion'' task
deal with
MIMO configurations,
where a set of original source signals
\ysssrcsigdisctimecentone\ to
\ysssrcsigdisctimecentlast\
are to be respectively 
restored on
the outputs
\yssoutsepsystsigdisctimecentone\
to
$
\yssoutsepsystsigdisctimecentvec
_
\Ysssrcnb
$
of the inverse system.

The blind MIMO version of the above system inversion problem is
almost the same as
blind source separation (BSS) 
{(see}
e.g.
{%
\ytextmodifartitihundredsixtyfourversiononestepthree{%
\cite{book-comon-jutten-ap,icabook-oja,amoi6-48,amansourieice2000}%
}%
}%
{)}%
:
As
in 
{system inversion,}
BSS 
aims at canceling 
the contributions of
all sources but one in each output signal of the separating 
system;
however, in BSS, one often 
allows each output
signal to be equal
to a source signal only up to an
acceptable
residual transform.
These transforms,
called indeterminacies,
cannot be avoided because only limited constraints
are set on the 
source signals 
and on the direct system 
which combines
(i.e., ``mixes'', in BSS terms) 
these signals.

Let us now consider
quantum 
``signals''
and systems,
where these ``signals''
e.g. consist of
the quantum states 
of
quantum bits, or qubits, defined below.
Then, among the above 
processing
problems,
the one 
which was
first
studied is non-blind 
system identification,
especially%
\footnote{%
See also
\cite{booknielsen}
p. 398 for the other earliest references.%
}
introduced in 1997 in 
\cite{amq30official}
and
often
called
``quantum process tomography''
(QPT)
by the quantum information processing 
(QIP)
community%
{: 
See%
}
e.g.
\cite{amq45official,amq50-physical-review,booknielsen,amq48,amq52-physical-review,amq-baldwin-physreva-2014,amq75,amq59,amq56,amq41}.
Besides, we introduced the field of
``quantum source separation'' (QSS) and especially its blind version
(BQSS)
in 2007%
{: 
See%
}
\cite{amoi5-31}.
We first mainly
developed a class of BQSS methods related to 
the
classical BSS methods based on
Independent Component Analysis:
See
especially
\ytextmodifartitihundredsixtyfourversiononestepthree{\cite{amoi6-18,amoi6-42}.}
We then proposed a second class of BQSS methods, based on
output
quantum state disentanglement:
See
especially
\cite{amoi6-34,amoi6-37,amoi6-64}.
Moreover,
in 
\ytextmodifartitihundredsixtyfourversiononestepthree{2015,}
we
introduced the field of
``blind quantum process tomography'' (BQPT) in
\cite{amoi6-46}.
We then 
\ytextmodifartitihundredsixtyfourversiononestepthree{developed it especially}
in
\cite{amoi6-118}.
We also
very recently 
\cite{amoi-arxiv-2021-sipqip}
introduced 
methods for a closely related problem,
namely
Blind Hamiltonian Parameter Estimation (BHPE).
All these QIP problems involve a quantum state transform,
to be identified or inverted.
Such a transform is also called a quantum process by the
QIP community,
or a quantum channel
\cite{booknielsen,preskill-http-chap3},
with
an explicit reference to the field of 
communications,
although the considered framework includes other
quantum application fields
in addition.

To solve 
the
above QIP problems, we first proposed
multiple-preparation methods, i.e. methods which require
many copies of each considered quantum state value,
in order to derive estimates of 
probabilities of associated measurement outcomes,
thus using statistical approaches,
as in usual QIP methods.
In addition, in
\ytextmodifartitihundredsixtyfourversiononestepthree{\cite{amoi6-104},}
we introduced the concept of
SIngle-Preparation QIP (or SIPQIP) methods,
i.e. methods 
that
can operate with only one instance of
each considered quantum state and 
that
extract information
thanks to statistical averaging over measurement
outcomes associated
with various states.
We then especially provided a detailed report of 
the principles
of this SIPQIP framework
and of its application 
to BQPT
in
\cite{amoi6-118}.
Finally, we very recently applied 
that framework to a variety of
QIP tasks:
See
\cite{amoi-arxiv-2021-sipqip}.
As shown further in this paper, this SIPQIP framework
is particularly
attractive for the communication applications considered here.

In multiple-source 
or multiple-qubit
configurations,
all above-defined classical or quantum processing
problems are expressed 
symmetrically with respect to all sources/qubits.
In contrast, other signal processing problems,
e.g.
related to communications, assign
a different role to different sources/qubits.
First considering classical communications again, 
this especially involves two-source configurations,
with
a ``main channel'' from,
say, Emitter 1 to Receiver 1 and with the
following two
possible cases.
In Case 0, the above configuration, from
Emitter 1, through the main channel, to Receiver 1,
does not interact with its environment.
In Case 1, such an interaction exists, i.e.
an ``intruder'',
often called the 
eavesdropper in the literature
\cite{booknielsen},
interferes in some way 
with the above channel from Emitter 1 to Receiver 1.
This yields the following two signal
processing problems.
On the one hand, 
the eavesdropper is interested
in extracting information
from the main channel.
This information is typically derived
by 
a 
Receiver 2
controlled by the eavesdropper,
who may in addition control 
an 
Emitter 2%
\ytextmodifartitihundredsixtyfourversiononestepthree{, that may be
considered as the jammer}
(this is the case in
the quantum extension of this scenario analyzed further
in this paper).
In contrast, 
the eavesdropper ideally requires no
prior knowledge about the
data manipulated by Emitter 1 and Receiver 1.
On the other hand,
Receiver 1 often has a strong interest in
intrusion detection, that is, in detecting
that the eavesdropper is extracting%
\ytextmodifartitihundredsixtyfourversiononestepthree{, i.e. intercepting,}
information
from the connection between Emitter 1 and
Receiver 1.
This receiver should perform this intrusion detection
task by using only the data that he receives from
the main channel (blind configuration) or by also
using information that he gets from Emitter 1
(non-blind mode; 
some configurations are also
stated to be semi-blind because Receiver 1 is
provided with \emph{very limited} prior information
in addition to the received data).
In all these configurations, Receiver 1 does not
access information about the data possibly sent by
Emitter 2 nor those obtained by Receiver 2.

How the above intrusion detection and
eavesdropping capabilities
extend to \emph{quantum} channels is currently
a major and 
still
open problem.
Whereas various approaches may be proposed to
this end, e.g. depending on the nature of the
considered quantum sources and channels,
this paper aims at investigating this topic
by analyzing how
results from
above-mentioned BQSS and 
related (i.e. BQPT and BHPE)
investigations reported so far,
hence with ``symmetric''
scenarios,
may be exploited to
derive 
first concepts for
intrusion detection and/or eavesdropping,
hence for
asymmetric scenarios.
We will thus 
only 
build upon
the 
data
model previously considered for BQSS and related tasks,
whereas the 
quantum processing
algorithms proposed in this paper 
are quite different from the above-mentioned BQSS 
and related algorithms,
since they have very different goals.
More precisely, we hereafter put the emphasis
on intrusion detection and 
more briefly
discuss 
eavesdropping.

The remainder of this paper is therefore
organized as follows.
We first present
the main features
of a single
qubit,
which is widely used as a building block for handling information
in QIP,
including in the problems tackled in this paper:
See
Section 
\ref{sec-onequbit}.
Then, in 
Section 
\ref{sec-def-two-qubits},
we summarize
the data model, i.e. the
considered class of channels
and its input, 
that was used in the above-mentioned BQSS and related methods,
when addressing coupling between
two qubits.
Both in Sections
\ref{sec-onequbit}
and
\ref{sec-def-two-qubits},
we also describe
measurements
associated
with the considered quantum states.
Moving to intrusion detection 
in
Section \ref{sec-intrusion},
new
data models
must first be derived from
the above one, because the information available
to Receiver 1
is not the same as in BQSS in this asymmetric scenario,
with or without interaction between both qubits.
Then,
two types of intrusion detection methods are proposed,
respectively using the
multiple-preparation and
single-preparation frameworks.
Finally, Section 
\ref{sec-concl}
contains a discussion of the features of the above
methods and of their extension to other hardware
implementations for practical
quantum communication
scenarios,
some 
considerations about the related
eavesdropping problem and a conclusion.

\section{Definition of a 
single
qubit}
\label{sec-onequbit}
Qubits
are 
widely
used 
instead of classical bits for performing computations
in the 
{field of 
QIP%
}
{\cite{booknielsen}}.
{Whereas
a classical bit can only take two values,
usually denoted 
as
0 and 1, 
at an initial
time
\yqubitonetimeinit\
a qubit
with index
\yqubitoneindexstd\
has a quantum state 
expressed, 
for a pure state, as
\begin{equation}
\label{eq-twoqubit-state-init-index-i}
| \psi_
\Yqubitoneindexstd
( \Yqubitonetimeinit )
{\rangle}
=
\alpha _
\Yqubitoneindexstd
| + 
{\rangle}
+ \beta _
\Yqubitoneindexstd
| - 
{\rangle}
\end{equation}
in the
basis defined by the two orthonormal vectors that we 
hereafter%
\footnote{These vectors
$
| + 
{\rangle}
$
and
$
| - 
{\rangle}
$
are 
often respectively denoted as
$|0
{\rangle}
$ and 
$|1
{\rangle}
$
(%
\ytextmodifartitihundredsixtyfourversiononestepthree{see}
e.g. \cite{booknielsen})%
,
especially when considering an abstract view of
qubits.
When having in mind the
physical implementation of qubits
as electron spins, as in 
most of
the present paper,
the notations 
$
| + 
{\rangle}
$
and
$
| - 
{\rangle}
$ 
are also widely used, with a reference to
spin component measurements along the quantization
axis, as detailed further in this paper.%
}
{denote
$
| + 
{\rangle}
$
and
$
| - 
{\rangle}
$},
where 
$ \alpha_
\Yqubitoneindexstd
$ and $ \beta_
\Yqubitoneindexstd
$
are two complex-valued
coefficients
constrained to meet the condition
\begin{equation}
\label{eq-onequbit-normalized-versionone}
| \alpha_
\Yqubitoneindexstd
| ^2 + | \beta_
\Yqubitoneindexstd
| ^2 = 1
\end{equation}
which expresses that
the state
$
| \psi 
_
\Yqubitoneindexstd
( \Yqubitonetimeinit )
{\rangle}
$
is normalized.}
In most of the literature,
$ 
\alpha_
\Yqubitoneindexstd
$ 
and 
$ 
\beta_
\Yqubitoneindexstd
$
are deterministic, i.e.
fixed, values
so that
$
| \psi_
\Yqubitoneindexstd
( \Yqubitonetimeinit )
\rangle
$
is a 
\emph{deterministic} pure state.
In part of our investigations dealing with BQSS and related tasks,
we also 
considered the case when
$ 
\alpha_
\Yqubitoneindexstd
$ 
and 
$ 
\beta_
\Yqubitoneindexstd
$
are random variables (RVs),
so that
$
| \psi_
\Yqubitoneindexstd
( \Yqubitonetimeinit )
\rangle
$
is a 
\emph{random} pure state
(%
See
e.g.
\cite{amoi6-18,amoi6-42,amoi6-67,amoi6-118}).

From a 
Quantum Physics (QP)
point of view, 
the above
abstract mathematical model
especially 
applies to
{electron 
spins 1/2,
which are quantum 
(i.e. non-classical) 
objects.%
}
The component of
{such a spin}%
,
with index
\yqubitoneindexstd,
along
a given arbitrary axis $Oz$ defines a
two-dimensional linear operator 
$
s_{
\Yqubitoneindexstd
z}.
$
The two eigenvalues of this
operator are equal to
$
+
\frac{1}{2}
$
and
$
-
\frac{1}{2}
$
in normalized units, and the corresponding eigenvectors are
therefore denoted
as
$
| + 
{\rangle}
$
and
$
| - 
{\rangle}
$.
The value obtained
when measuring this spin component
can only be
$
+
\frac{1}{2}
$
or
$
-
\frac{1}{2}
$.
Moreover,
let us assume
this spin is in the state
$
| \psi 
_{
\Yqubitoneindexstd
}
( \Yqubitonetimeinit )
{\rangle}
$
defined by
(\ref{eq-twoqubit-state-init-index-i})
when
performing such a
measurement.
Then, the probability that the measured value is equal to
$
+
\frac{1}{2}
$
(respectively
$
-
\frac{1}{2}
$)
is equal to
$
| \alpha 
_{
\Yqubitoneindexstd
}
| ^2
$
(respectively
$
| \beta 
_{
\Yqubitoneindexstd
}
| ^2
$),
i.e. to the squared modulus of the coefficient in
(\ref{eq-twoqubit-state-init-index-i})
of the associated eigenvector
$
| + 
{\rangle}
$
(respectively
$
| - 
{\rangle}
$).

The above discussion concerns the state of the considered spin at
a given 
initial time
\yqubitonetimeinit .
This state then
evolves with time.
The
spin is here supposed to be
placed in a
{static}
magnetic field
and thus
coupled to it.
The time interval when it is
considered
is
assumed to be short
enough for the coupling
between the spin and its environment
to be
negligible.
In these conditions, the spin
has a Hamiltonian
\ytextmodifartitihundredsixtyfourversiononestepthree{\cite{booknielsen}.}
Therefore, if the spin state
$
| \psi 
_{
\Yqubitoneindexstd
}
( t_0) 
{\rangle}
$
at time
$
t_0 $
is defined by
(\ref{eq-twoqubit-state-init-index-i}),
it then
evolves according to
Schr\" odinger's
equation
and its value at any 
subsequent
time $t$
is
\begin{equation}
\label{eq-state-onequbit-anytime}
| \psi 
_{
\Yqubitoneindexstd
}
(t) 
{\rangle}
=
\alpha
_{
\Yqubitoneindexstd
}
e^{-
\Ysqrtminusone
\omega _p
( t - t_0 )
}
| + 
{\rangle}
+ \beta
_{
\Yqubitoneindexstd
}
e^{-
\Ysqrtminusone
\omega _m
( t - t_0 )
}
| - 
{\rangle}
\end{equation}
where 
the real (angular) frequencies
$
\omega _p
$
and
$
\omega _m
$
depend on the considered physical setup
\ytextmodifartitihundredsixtyfourversiononestepthree{and
\ysqrtminusone\
is the imaginary unit.}

\section{%
Considered quantum channels:
Coupling
model for two qubits}
\label{sec-def-two-qubits}
The above description directly applies to several qubits if they are not ``coupled",
i.e. if they do 
{not}
interact with one another.
However,
coupling between individual 
quantum states
has
to be considered in the QIP/QP
area,
in the
same way 
{%
{as}
signal coupling  
exists
in various
\emph{classical} signal processing systems.}
Coupling in quantum physical setups
e.g. occurs when two electron spins interact 
{through exchange}.
In 
\cite{amoi6-18},
we considered
a
two-qubit
system composed of
two 
distinguishable
\cite{amoi6-64}
spins 
coupled
according to
the version of the Heisenberg model which has a
cylindrical-symmetry axis, denoted $Oz$
{and collinear to the applied magnetic field.}
We analyzed in detail
the global state 
of that two-qubit system
resulting from that coupling
and 
the associated measured values.
Here again, the measured value of the
component of
each spin
along
axis $Oz$ 
can only be
$
+
\frac{1}{2}
$
or
$
-
\frac{1}{2}
$.
Therefore, when measuring the components of both spins, the obtained couple
of values is 
equal 
to one of the four possible values
$(+\frac{1}{2},+\frac{1}{2})$,
$(+\frac{1}{2},-\frac{1}{2})$,
$(-\frac{1}{2},+\frac{1}{2} )$
and
$(-\frac{1}{2},-\frac{1}{2})$.
The probabilities of these four values are respectively denoted
as
\ytwoqubitsprobaplusplus , 
\ytwoqubitsprobaplusminus , \ytwoqubitsprobaminusplus\
and
\ytwoqubitsprobaminusminus\
hereafter.
These probabilities are related as follows to
the state
of the overall system composed of these two
spins.
This state may be expressed as a linear combination of the
vectors of
the four-dimensional basis
$
\{ | ++ 
{\rangle}
, | +- 
{\rangle}
, | -+ 
\ytextmodifartitihundredsixtyfourversiononestepthree{\rangle}
, | -- 
{\rangle}
\}
$
which corresponds to the operators $s_{1z}$ and $s_{2z}$ respectively
associated 
with
the components
of 
Spin 1 and Spin 2
along
the symmetry axis $Oz$.
As in Section
\ref{sec-onequbit},
each of the 
probabilities 
\ytwoqubitsprobaplusplus\ to
\ytwoqubitsprobaminusminus\
is here
equal to the squared modulus
of the coefficient of the corresponding basis vector in the expression of the overall
system state.
In 
\cite{amoi6-18},
we provided a detailed derivation of the expressions of these probabilities
in the following configuration.
The two spins
are separately
initialized 
(i.e. prepared)
at time
$
t_0 $,
with states 
$
| \psi_
{
\Yqubitoneindexstd
}
( \Yqubitonetimeinit ) \rangle
$
defined by
(\ref{eq-twoqubit-state-init-index-i})%
,
where
$
\Yqubitoneindexstd
=
1
$
for  
Spin
1 
and
$
\Yqubitoneindexstd
=
2
$
for 
Spin
2,
first considering deterministic pure states.
The 
initial
state
of the overall system composed of these two
distinguishable
spins
is therefore
equal to the tensor product 
(denoted as
$
\otimes
$)
of the
states
of both spins
defined in
(\ref{eq-twoqubit-state-init-index-i}),
i.e.
\begin{equation}
\label{eq-qubitbothtimeinitstate-tensor-prod}
\Ymixsyststateinitial
=
\Yqubitonetimeinitstateindexone
\otimes
\Yqubitonetimeinitstateindextwo
.
\end{equation}
That initial state 
\ymixsyststateinitial\
is thus
unentangled.
The 
overall
system state then evolves with time
and 
the spin
states thus
get ``mixed" 
(in the 
classical
BSS
sense%
\footnote{%
The terms ``mixing'' and ``mixtures'' should be 
considered
with care when
dealing with
quantum data: 
\ytextmodifartitihundredsixtyfourversiononesteptwo{%
In
this paper, 
when speaking of random pure states, we implicitly refer to some 
statistical mixtures, as defined in quantum mechanics, but, except 
in the present note, we do not explicitly use the 
expression 
``statistical mixture''.}%
}%
) 
with one another%
,
thus yielding an entangled state
\ymixsyststatefinal\
(except for very specific parameter values).
The time evolution of the overall system state
is defined by
phase rotations, 
as in
(\ref{eq-state-onequbit-anytime}),
and this here involves
four 
frequencies.
These frequencies 
depend on
\ytextmodifartitihundredsixtyfourversiononestepthree{the}
Heisenberg coupling,
which is especially
characterized by
the so-called principal value 
$J_{xy}$ 
of the exchange tensor
(see
\cite{amoi6-18}
for more details).
We derived the expressions of the above probabilities
\ytwoqubitsprobaplusplus\
to
\ytwoqubitsprobaminusminus\
at an arbitrary
time
$
t > t_0 $,
with respect
to the polar representation of the 
initial
qubit parameters
$ \alpha
_{
\Yqubitoneindexstd
}
$ and $ \beta
_{
\Yqubitoneindexstd
}
$,
which reads
\begin{equation}
\label{eq-def-qubit-polar-qubit-indexstd} 
\alpha
_{
\Yqubitoneindexstd
}
=
\Yparamqubitindexstdstateplusmodulus
e ^{ \Ysqrtminusone
\Yparamqubitindexstdstateplusphase }
\hspace{5mm} 
\beta 
_{
\Yqubitoneindexstd
}
=
\Yparamqubitindexstdstateminusmodulus e ^{ \Ysqrtminusone
\Yparamqubitindexstdstateminusphase }
\hspace{10mm} 
\Yqubitindexstd \in \{ 1 , 2 \}
\end{equation}
with
$
0
\leq
\Yparamqubitindexstdstateplusmodulus
\leq
1
$
and
\begin{equation}
\Yparamqubitindexstdstateminusmodulus
=
\sqrt
{
1
-
\Yparamqubitindexstdstateplusmodulus
^2
}
\label{eq-paramqubitindexstdstateminusmodulus-vs-paramqubitindexstdstateplusmodulus}
\end{equation}
due to
(\ref{eq-onequbit-normalized-versionone}).
The above probabilities may then be expressed as follows:
\begin{eqnarray}
\label{eq-statecoef-vs-twoqubitsprobaplusplus-polar}
\Ytwoqubitsprobaplusplus
&
=
& 
\Yparamqubitonestateplusmodulus ^2 \Yparamqubittwostateplusmodulus
^2 
\\
\nonumber
\Ytwoqubitsprobaplusminus 
&
=
&
\Yparamqubitonestateplusmodulus ^2 ( 1
-
\Yparamqubittwostateplusmodulus ^2 ) ( 1 -
\Ytwoqubitresultphaseevolsin ^2 ) + ( 1
-
\Yparamqubitonestateplusmodulus ^2 )
\Yparamqubittwostateplusmodulus ^2 \Ytwoqubitresultphaseevolsin ^2
\\
&
&
{
-
2 \Yparamqubitonestateplusmodulus \Yparamqubittwostateplusmodulus
\sqrt{ 1
-
\Yparamqubitonestateplusmodulus ^2 } \sqrt{ 1
-
\Yparamqubittwostateplusmodulus ^2 } \sqrt{ 1 -
\Ytwoqubitresultphaseevolsin ^2 } \Ytwoqubitresultphaseevolsin
\sin \Ytwoqubitresultphaseinit
}
\label{eq-statecoef-vs-twoqubitsprobaplusminus-polar-versionthree}
\\
\label{eq-statecoef-vs-twoqubitsprobaminusminus-polar-vs-paramqubitindexstdstateplusmodulus}
\Ytwoqubitsprobaminusminus 
&
=
&
( 1
-
\Yparamqubitonestateplusmodulus ^2 ) ( 1
-
\Yparamqubittwostateplusmodulus ^2 )
\end{eqnarray}
with
\begin{eqnarray}
\label{eq-def-twoqubitresultphaseinit-twoqubitresultphaseevol}
\Ytwoqubitresultphaseinit
&
=
&
( \Yparamqubittwostateminusphase
-
\Yparamqubittwostateplusphase
)
-
( 
\Yparamqubitonestateminusphase
-
\Yparamqubitonestateplusphase )
\\
\label{eq-def-Ytwoqubitresultphaseevol-versiontwo}
\Ytwoqubitresultphaseevol
&
=
&
- \frac{ J_{xy}
( t - t_0 )
} { \hbar } 
\\
\Ytwoqubitresultphaseevolsin
&
=
&
\mbox{sgn} ( \cos \Ytwoqubitresultphaseevol )
\sin \Ytwoqubitresultphaseevol
\label{eq-def-Ytwoqubitresultphaseevol-versionthree}
\end{eqnarray}
where
$
\hbar
$
is the reduced Planck constant.
Probability
\ytwoqubitsprobaminusplus\ 
{is not
considered, since 
it may be derived from the other three 
\ytextmodifartitihundredsixtyfourversiononestepthree{probabilities}
by means of%
}
\begin{equation}
\label{eq-twobits-sum-proba-equal-one} \Ytwoqubitsprobaplusplus +
\Ytwoqubitsprobaplusminus + \Ytwoqubitsprobaminusplus +
\Ytwoqubitsprobaminusminus
=
1
.
\end{equation}

Eq.
(\ref{eq-statecoef-vs-twoqubitsprobaplusplus-polar})-%
(\ref{eq-statecoef-vs-twoqubitsprobaminusminus-polar-vs-paramqubitindexstdstateplusmodulus})
yield a QSS problem because, using the 
classical BSS
terminology, 
{they show}
that some
``observations" are ``mixtures" of the 
quantities which define
quantum ``sources".
This ``mixing model"
(\ref{eq-statecoef-vs-twoqubitsprobaplusplus-polar})-%
(\ref{eq-statecoef-vs-twoqubitsprobaminusminus-polar-vs-paramqubitindexstdstateplusmodulus})
involves the following items.
The observations 
are the probabilities
\ytwoqubitsprobaplusplus , 
\ytwoqubitsprobaplusminus\
and
\ytwoqubitsprobaminusminus\
measured 
for each 
choice of the initial states 
(\ref{eq-twoqubit-state-init-index-i})
of the qubits.
More precisely, these probabilities are not known 
exactly but estimated in practice.
The procedure that we 
used
to this end 
e.g.
in 
\cite{amoi6-18,amoi6-42}%
, and that is also widely employed in the QIP literature
\cite{amq75,amq30official},
operates as follows
for each 
choice of the initial states 
(\ref{eq-twoqubit-state-init-index-i})
of the qubits.
We repeatedly perform two operations: 
i) we first initialize these qubits 
according
to
(\ref{eq-twoqubit-state-init-index-i})
and ii)
after a fixed time interval
when 
coupling occurs,
we measure the two
spin components 
along $Oz$
associated 
with
the system composed of
these two coupled qubits.
The relative frequencies of 
occurrence
of all four possible couples of values of spin components
(i.e.
$(+\frac{1}{2},+\frac{1}{2})$
to 
$(-\frac{1}{2},-\frac{1}{2})$)
then yield estimates of the corresponding probabilities.
This approach therefore
requires 
a large number
(typically from a few 
{thousand}
up to 
a few
hundred 
{thousand}
\cite{amoi6-18,amoi6-64})
of copies of 
the considered two-qubit
state.
At this stage,
we ignore
the resulting estimation errors and therefore consider
the exact mixing model
(\ref{eq-statecoef-vs-twoqubitsprobaplusplus-polar})-%
(\ref{eq-statecoef-vs-twoqubitsprobaminusminus-polar-vs-paramqubitindexstdstateplusmodulus}).
Using standard 
BSS
notations, the 
observation vector
is therefore
$
\Yssmixsigdisctimecentvec
=
[ \Yssmixsigdisctimecentone , \Yssmixsigdisctimecenttwo ,
\Yssmixsigdisctimecentthree
 ] ^{T}
$%
, where
$
^T
$
stands for transpose and%
\footnote{%
{It should be noted that
the observed signals involved in this QSS problem have a specific nature,
as compared 
with
standard
non-quantum
BSS problems.
In the latter 
problems,
each value of an observed signal 
is usually the value of a measured physical quantity, such as the
value of a voltage measured at a given time.
In contrast,
as shown by
(\ref{eq-artiti7v1-observ-not-std-vs-qss}),
each value of an observed signal 
is here the value of a \emph{probability}
(which is estimated in practice).
The overall signal composed of all successive values of a given
observation
(e.g. all values of
\yssmixsigdisctimecentone )
therefore consists of a set of values of
probabilities
(e.g. all values of
\ytwoqubitsprobaplusplus ),
which depend on
the values of the 
\ytextmodifartitihundredsixtyfourversiononesteptwo{coefficients
used for initializing the qubit states.}%
}}
\begin{equation}
\label{eq-artiti7v1-observ-not-std-vs-qss}
\Yssmixsigdisctimecentone = \Ytwoqubitsprobaplusplus,
\hspace{5mm}
\Yssmixsigdisctimecenttwo = \Ytwoqubitsprobaplusminus
,
\hspace{5mm}
\Yssmixsigdisctimecentthree = \Ytwoqubitsprobaminusminus
.
\end{equation}
{Eq. 
(\ref{eq-statecoef-vs-twoqubitsprobaplusplus-polar})-%
(\ref{eq-statecoef-vs-twoqubitsprobaminusminus-polar-vs-paramqubitindexstdstateplusmodulus}) show that the}
source vector
to be retrieved 
from these observations
{turns out to be}
$
\Ysssrcsigdisctimecentvec
=
[ \Ysssrcsigdisctimecentone , \Ysssrcsigdisctimecenttwo ,
\Ysssrcsigdisctimecentthree
 ] ^{T}
$
with
$
\Ysssrcsigdisctimecentone = \Yparamqubitonestateplusmodulus,
\Ysssrcsigdisctimecenttwo = \Yparamqubittwostateplusmodulus
$
and $ \Ysssrcsigdisctimecentthree = \Ytwoqubitresultphaseinit $.
The parameters
\yparamqubitindexstdstateminusmodulus\
are then 
derived 
from
(\ref{eq-paramqubitindexstdstateminusmodulus-vs-paramqubitindexstdstateplusmodulus}).
The four
phase parameters in
(\ref{eq-def-qubit-polar-qubit-indexstd})
cannot be individually extracted
from
their combination 
\ytwoqubitresultphaseinit\
(%
anyway,
only 
the phase differences
$
(
\Yparamqubitindexstdstateminusphase
-
\Yparamqubitindexstdstateplusphase
)
$
have a physical
meaning
\cite{amoi6-118}%
).
The transform from the sources to the observations defined by the 
\ytextmodifartitihundredsixtyfourversiononesteptwo{non-linear}
mixing model
(\ref{eq-statecoef-vs-twoqubitsprobaplusplus-polar})-%
(\ref{eq-statecoef-vs-twoqubitsprobaminusminus-polar-vs-paramqubitindexstdstateplusmodulus})
involves a single ``mixing parameter", 
{namely}
\ytwoqubitresultphaseevolsin .
As shown by
(\ref{eq-def-Ytwoqubitresultphaseevol-versionthree}),
this parameter 
always meets the condition
$
0 
\leq
\Ytwoqubitresultphaseevolsin ^2
\leq
1
$.
In most
configurations,
the values of the coupling parameter
$
J_{xy}
$
and therefore of
\ytwoqubitresultphaseevolsin\
(see
(\ref{eq-def-Ytwoqubitresultphaseevol-versiontwo})-%
(\ref{eq-def-Ytwoqubitresultphaseevol-versionthree}))
are unknown 
(the sign of
$
J_{xy}
$
is 
{however known
in some configurations%
}%
)
\ytextmodifartitihundredsixtyfourversiononestepthree{but fixed, i.e. deterministic}%
.
This corresponds to the \emph{blind} version of this
QSS problem.
In this configuration,
estimating
the sources first requires one to
estimate the
unknown
mixing parameter
\ytwoqubitresultphaseevolsin .
BQSS is thus an estimation problem
\cite{book-scharf},
where one aims at deriving
continuous-valued quantities.
In contrast, as shown below,
intrusion detection
is a decision making (i.e. detection) problem
\cite{book-scharf},
with a ``yes/no answer''
to the determination of which case, among two possible cases, is
actually faced.
The algorithms proposed below to answer that decision problem are therefore
quite different from those
\ytextmodifartitihundredsixtyfourversiononestepthree{introduced}
in our previous papers to
solve the BQSS problem.

\section{%
Intrusion detection}
\label{sec-intrusion}
\subsection{%
Data models for intrusion detection}
The BQSS methods that we proposed in our previous papers
to address the data model of Section
\ref{sec-def-two-qubits}
take advantage of all the data that are available in
that model, that is, of the
probabilities
\ytwoqubitsprobaplusplus , 
\ytwoqubitsprobaplusminus\
and
\ytwoqubitsprobaminusminus\
that are derived (estimated in practice)
from spin component measurements associated with \emph{both}
qubits. This corresponds to the above-defined symmetric
scenario.
In contrast, these 
``two-qubit probabilities'', i.e. 
joint probabilities,
are not known any more in
the investigation reported in the present paper, due to
the considered asymmetric scenario.
More precisely, if 
starting from
the data model of Section
\ref{sec-def-two-qubits}
as a toy model for quantum channels at this stage,
the considered intrusion detection scenario may be defined
as follows.
The main channel 
considered
in Section
\ref{sec-intro}
here goes from the 
initial 
deterministic
pure
state
\yqubitonetimeinitstateindexone\
involved in
(\ref{eq-qubitbothtimeinitstate-tensor-prod}),
provided by Emitter 1,
to the results of the measurements performed
by Receiver 1
at the final time
\yqubitonetimefinal\
for the \emph{first} qubit of the data model
of
Section
\ref{sec-def-two-qubits}%
.
These measurements
thus only allow
Receiver 1 to access ``one-qubit probabilities'', i.e. marginal
probabilities, associated with 
Qubit 1
(at the final time
\yqubitonetimefinal ).

When intrusion is actually performed,
the above 
model also involves coupling with the
\emph{second} qubit
(see Case 1 in Section
\ref{sec-intro}).
The probabilities of measurement results of Receiver 1 thus also
depend 
(i)
on
the initial 
state
\yqubitonetimeinitstateindextwo\
involved in
(\ref{eq-qubitbothtimeinitstate-tensor-prod})
and
provided by Emitter 2,
and
(ii) on the qubit
coupling phenomenon leading to
(\ref{eq-statecoef-vs-twoqubitsprobaplusplus-polar})-%
(\ref{eq-statecoef-vs-twoqubitsprobaminusminus-polar-vs-paramqubitindexstdstateplusmodulus})
and hence on the parameter
\yexchangetensorppalvaluexy\
of that exchange coupling model. 
We then aim at defining and exploiting the 
probabilities of the measurement results of
Receiver 1.
This 
may be performed
as follows,
still considering
the data model of Section
\ref{sec-def-two-qubits}.
At the final 
time
\yqubitonetimefinal,
the measurements for each qubit with index
$
\Yqubitindexstd
\in
\{
1,
2
\}
$
define a 
binary
RV 
denoted as
\ymixmodelstateoutqubitindexstddirectionnonerv ,
whose
possible values are equal to
$
+
\frac{1}{2} 
$
and 
$
-
\frac{1}{2} 
$.
The two events defined by the outcomes of this RV are therefore
denoted as
$
\{
\Ymixmodelstateoutqubitindexstddirectionnonerv
=
+
\}
$
and
$
\{
\Ymixmodelstateoutqubitindexstddirectionnonerv
=
-
\}
$
hereafter.
The joint probabilities of the two RVs
defined by the considered two qubits,
namely
$
\Yprob
(
\Ymixmodelstateoutqubitindexonedirectionnonerv
=
+
,
\Ymixmodelstateoutqubitindextwodirectionnonerv
=
+
)
$,
$
\Yprob
(
\Ymixmodelstateoutqubitindexonedirectionnonerv
=
+
,
\Ymixmodelstateoutqubitindextwodirectionnonerv
=
-
)
$,
$
\Yprob
(
\Ymixmodelstateoutqubitindexonedirectionnonerv
=
-
,
\Ymixmodelstateoutqubitindextwodirectionnonerv
=
+
)
$
and
$
\Yprob
(
\Ymixmodelstateoutqubitindexonedirectionnonerv
=
-
,
\Ymixmodelstateoutqubitindextwodirectionnonerv
=
-
)
$
are nothing but the above-defined probabilities
\ytwoqubitsprobaplusplus , 
\ytwoqubitsprobaplusminus ,
\ytwoqubitsprobaminusplus\
and
\ytwoqubitsprobaminusminus .
Besides, the marginal probabilities associated with
measurements performed for Qubit 1 only may be
expressed as
follows
\begin{eqnarray}
\Yprob
(
\Ymixmodelstateoutqubitindexonedirectionnonerv
=
+
)
&
=
&
\Yprob
(
\Ymixmodelstateoutqubitindexonedirectionnonerv
=
+
,
\Ymixmodelstateoutqubitindextwodirectionnonerv
=
+
)
+
\Yprob
(
\Ymixmodelstateoutqubitindexonedirectionnonerv
=
+
,
\Ymixmodelstateoutqubitindextwodirectionnonerv
=
-
)
\nonumber
\\
&
&
\\
&
=
&
\Ytwoqubitsprobaplusplus
+
\Ytwoqubitsprobaplusminus
.
\label{eq-mixmodelstateoutqubitindexonedirectionnonerv-prob-plus-vs-twoqubitsprobaplusplus-twoqubitsprobaplusminus}
\end{eqnarray}
Using
(\ref{eq-statecoef-vs-twoqubitsprobaplusplus-polar})-%
(\ref{eq-statecoef-vs-twoqubitsprobaplusminus-polar-versionthree}),
this yields
\begin{eqnarray}
\Yprob
(
\Ymixmodelstateoutqubitindexonedirectionnonerv
=
+
)
&
=
&
\Yparamqubitonestateplusmodulus ^2 
+
\Ytwoqubitresultphaseevolsin ^2
(
\Yparamqubittwostateplusmodulus ^2
-
\Yparamqubitonestateplusmodulus ^2 
)
\nonumber
\\
&
&
{
-
2 \Yparamqubitonestateplusmodulus \Yparamqubittwostateplusmodulus
\sqrt{ 1
-
\Yparamqubitonestateplusmodulus ^2 } \sqrt{ 1
-
\Yparamqubittwostateplusmodulus ^2 } \sqrt{ 1 -
\Ytwoqubitresultphaseevolsin ^2 } \Ytwoqubitresultphaseevolsin
\sin \Ytwoqubitresultphaseinit
}
.
\nonumber
\\
&
&
\label{eq-mixmodelstateoutqubitindexonedirectionnonerv-proba-plus-with-intrusion}
\end{eqnarray}
Besides,
$
\Yprob
(
\Ymixmodelstateoutqubitindexonedirectionnonerv
=
-
)
$
provides no additional information, because
\begin{equation}
\Yprob
(
\Ymixmodelstateoutqubitindexonedirectionnonerv
=
-
)
=
1
-
\Yprob
(
\Ymixmodelstateoutqubitindexonedirectionnonerv
=
+
)
.
\end{equation}

As stated above, this model corresponds to the case
when intrusion actually occurs (Case 1),
which physically corresponds to electrons being close to one
another
\ytextmodifartitihundredsixtyfourversiononesteptwo{or both close to the 
same atom/ion}%
, hence with exchange coupling involving
$
\Yexchangetensorppalvaluexy
\neq
0
$.
Let us now consider the case when no intrusion is
performed (Case 0), which physically corresponds to electrons being 
far from one
another. 
The corresponding data model may be derived by setting
$
\Yexchangetensorppalvaluexy
=
0
$
in the data model
defined above for Case 1.
Eq.
(\ref{eq-def-Ytwoqubitresultphaseevol-versiontwo})-%
(\ref{eq-def-Ytwoqubitresultphaseevol-versionthree})
then yield
$
\Ytwoqubitresultphaseevolsin
=
0
$,
so that
(\ref{eq-mixmodelstateoutqubitindexonedirectionnonerv-proba-plus-with-intrusion})
reduces to
\begin{equation}
\Yprob
(
\Ymixmodelstateoutqubitindexonedirectionnonerv
=
+
)
=
\Yparamqubitonestateplusmodulus ^2 
.
\label{eq-mixmodelstateoutqubitindexonedirectionnonerv-proba-plus-no-intrusion}
\end{equation}
This result could be anticipated as follows. We here consider the case
when Qubit 1 does not interact with Qubit 2, so that its state evolves
according to
(\ref{eq-state-onequbit-anytime})
with
$
\Yqubitoneindexstd
=
1
$.
Therefore, as explained above, at any time
\yqubitonetimefinal,
the probability of
$
\{
\Ymixmodelstateoutqubitindexonedirectionnonerv
=
+
\}
$
is equal to
the squared modulus of the coefficient in
(\ref{eq-state-onequbit-anytime})
of the vector
$
| + 
\rangle
$.
It is thus equal to
$
|
\alpha
_{
1
}|
^2
$,
i.e.
$
\Yparamqubitonestateplusmodulus ^2 
$
due to
(\ref{eq-def-qubit-polar-qubit-indexstd}).
\subsection{Multiple-preparation intrusion detection methods}
\label{sec-intrusion-multiple-prep}
The problem addressed in this paper consists of only using
the measurements performed by Receiver 1
so as to determine whether intrusion occurs or not,
i.e. whether the main channel is in Case 0 or Case 1.
That can be seen as a hypothesis testing problem
(i.e. a decision making or detection
problem)
\cite{book-scharf}, with hypotheses H0 and H1 respectively
corresponding to the above-defined Cases 0 and 1.

To perform the above test, several methods may be proposed.
Their simplest version
uses a single,
deterministic,
value of the initial
states
$
| \psi_
{
\Yqubitoneindexstd
}
(
\Yqubitonetimeinit
)
\rangle
$
with
$
\Yqubitoneindexstd
\in
\{
1
,
2
\}
$,
of the final state
\ymixsyststatefinal\
of the two-qubit system at time
\yqubitonetimefinal\
and of the associated probability
$
\Yprob
(
\Ymixmodelstateoutqubitindexonedirectionnonerv
=
+
)
$.
This
method exploits the fact that,
``in general'',
this probability does not take the same value
depending whether Case 0 or Case 1 is considered, as shown
by
(\ref{eq-mixmodelstateoutqubitindexonedirectionnonerv-proba-plus-with-intrusion})
and
(\ref{eq-mixmodelstateoutqubitindexonedirectionnonerv-proba-plus-no-intrusion}).
By ``in general'',
we mean that the values in
(\ref{eq-mixmodelstateoutqubitindexonedirectionnonerv-proba-plus-with-intrusion})
and
(\ref{eq-mixmodelstateoutqubitindexonedirectionnonerv-proba-plus-no-intrusion})
are different except for very specific values of
the quantities that they involve, that are related to the
initial quantum state
(parameters $
\Yparamqubitonestateplusmodulus
,
\Yparamqubittwostateplusmodulus
,
\Ytwoqubitresultphaseinit
$)
or to the channel
(parameters
\ytwoqubitresultphaseevolsin\
and hence
\yexchangetensorppalvaluexy\
and
$
(
\Yqubitonetimefinal
-
\Yqubitonetimeinit
)
$).
We here ignore these very specific cases 
and
we will further discuss 
this topic in Section
\ref{sec-concl}.
One should also keep in mind that,
in practice, an estimate of
$
\Yprob
(
\Ymixmodelstateoutqubitindexonedirectionnonerv
=
+
)
$
is used and, to obtain it, one 
must prepare \emph{many} copies
of the initial state 
\ymixsyststateinitial\
of the two-qubit system, as explained in
Section 
\ref{sec-def-two-qubits}.

A first intrusion detection method 
then
consists of using
a value
\yqubitonetimeinitstateindexone\
of the state 
provided by Emitter 1,
or at least a value of its parameter
$
\Yparamqubitonestateplusmodulus
$
in
(\ref{eq-def-qubit-polar-qubit-indexstd}),
that Receiver 1 knows.
Receiver 1 then compares
his estimate of
$
\Yprob
(
\Ymixmodelstateoutqubitindexonedirectionnonerv
=
+
)
$
to
$
\Yparamqubitonestateplusmodulus ^2 
$
and 
makes the following decision,
based on
(\ref{eq-mixmodelstateoutqubitindexonedirectionnonerv-proba-plus-with-intrusion})
and
(\ref{eq-mixmodelstateoutqubitindexonedirectionnonerv-proba-plus-no-intrusion}):
If
that estimate of
$
\Yprob
(
\Ymixmodelstateoutqubitindexonedirectionnonerv
=
+
)
$
is ``close enough''
(one may aim at deriving a bound
from test theory for given
statistics of the considered data)
to
$
\Yparamqubitonestateplusmodulus ^2 
$,
then
Receiver 1 decides that no intrusion is being
performed (Case 0); otherwise, he decides that
intrusion is occurring (Case 1).
Since
this approach requires Receiver 1 to 
know a 
value of emitted (i.e. source) data,
it can be considered to be a non-blind method.
Of course, such an approach can only be used 
to perform detection
intrusion during one or a few limited time periods%
\ytextmodifartitihundredsixtyfourversiononestepthree{, moreover}
jointly defined by Emitter 1 and Receiver 
\ytextmodifartitihundredsixtyfourversiononestepthree{1 (somewhat as when using
a synchronization sequence
 in classical communication networks)}
because, otherwise,
Receiver 1 would have to permanently know which states
are provided by Emitter 1, which would make data
transmission in the main channel useless.

To reduce 
the above restriction
about known
emitted data, one may instead
develop a blind variant of the above method,
i.e. a variant in which Receiver 1 does not know which
state is provided by Emitter 1 
(and by Emitter 2)
and can only use estimates of
$
\Yprob
(
\Ymixmodelstateoutqubitindexonedirectionnonerv
=
+
)
$.
The proposed approach then consists of splitting the
above-mentioned complete set of copies of the
initial state
\ymixsyststateinitial\
in two successive subsets.
An estimate of
$
\Yprob
(
\Ymixmodelstateoutqubitindexonedirectionnonerv
=
+
)
$
is then separately computed by Receiver 1 for each
subset and these two estimates are then compared:
If
they are ``far enough'' (with the same comment
as above concerning an associated bound)
from one another, 
Receiver 1
considers 
that the main channel switched between
Cases 0 and 1 
(or between Case 1 with one value of
\ytwoqubitresultphaseevolsin\
to Case 1 with another value)
from one of the above subsets to the other.

Both variants of this method have limitations.
In particular, they require Receiver 1
to estimate at least one
value of the probability
$
\Yprob
(
\Ymixmodelstateoutqubitindexonedirectionnonerv
=
+
)
$,
which requires many copies of the same state 
\yqubitonetimeinitstateindexone\
to be transmitted by Emitter 1
through the main channel 
\ytextmodifartitihundredsixtyfourversiononestepthree{(with a timing known
by Receiver 1)}
and, more importantly, 
many copies of the
same state 
\yqubitonetimeinitstateindextwo\
to be provided 
by Emitter 2 
(i.e. the intruder)
meanwhile,
which is
very constraining from a practical point of view.
If sticking to 
that multi-preparation framework that is usual in QIP,
there might seem to be no 
\ytextmodifartitihundredsixtyfourversiononesteptwo{solution}
to this problem at first glance,
because estimating a probability value requires a large number
of trials for the considered single experiment.
However, beyond that usual QIP framework, we recently
developed an original concept
(see
\cite{amoi6-104,amoi6-118,amoi-arxiv-2021-sipqip}),
called SIngle-Preparation QIP (or SIPQIP) for its general
version, and especially applied to BQSS and related tasks
so far, 
which
solves the above problem, as will now be shown.
Briefly,
instead of 
estimating a single 
deterministic probability
from many copies of a single deterministic quantum state,
SIPQIP
estimates the expectation of a random probability
associated with 
a random quantum 
\ytextmodifartitihundredsixtyfourversiononesteptwo{pure}
state,
i.e. associated with various quantum states 
\ytextmodifartitihundredsixtyfourversiononesteptwo{whose coefficients
(such as
$
\alpha _
\Yqubitoneindexstd
$
and
$
\beta _
\Yqubitoneindexstd
$
in
(\ref{eq-twoqubit-state-init-index-i}))
are}
randomly drawn
in practice,
by possibly using a single instance of each of these states.
\subsection{Single-preparation intrusion detection methods}
We here address
the situation when 
Receiver 1 considers
the initial state
(\ref{eq-qubitbothtimeinitstate-tensor-prod})
of the two-qubit system, and hence 
the initial state
(\ref{eq-twoqubit-state-init-index-i})
of each qubit,
as a \emph{random} pure state,
i.e. when the parameters
\yparamqubitindexstdstateplusmodulus\
(and hence
\yparamqubitindexstdstateminusmodulus ),
\yparamqubitindexstdstateplusphase\
and
\yparamqubitindexstdstateminusphase\
in
(\ref{eq-def-qubit-polar-qubit-indexstd})
are RVs
\ytextmodifartitihundredsixtyfourversiononesteptwo{(this concept of
random quantum pure state is defined in more detail in
\cite{amoi6-67})}%
.
Then, the
probabilities
\ytwoqubitsprobaplusplus , 
\ytwoqubitsprobaplusminus\
and
\ytwoqubitsprobaminusminus\
in
(\ref{eq-statecoef-vs-twoqubitsprobaplusplus-polar})-%
(\ref{eq-statecoef-vs-twoqubitsprobaminusminus-polar-vs-paramqubitindexstdstateplusmodulus})
are also
RVs,
and so is the probability
$
\Yprob
(
\Ymixmodelstateoutqubitindexonedirectionnonerv
=
+
)
$.
The approach proposed here is then based on estimating the
\emph{expectation} of
$
\Yprob
(
\Ymixmodelstateoutqubitindexonedirectionnonerv
=
+
)
$
over random states
(\ref{eq-twoqubit-state-init-index-i}).
Due to
(\ref{eq-mixmodelstateoutqubitindexonedirectionnonerv-prob-plus-vs-twoqubitsprobaplusplus-twoqubitsprobaplusminus}),
this expectation
may be expressed with respect to the expectations
of
\ytwoqubitsprobaplusplus\
and
\ytwoqubitsprobaplusminus .
A major property is then that all
these expectations may in practice be
estimated 
by using only one instance of each of the considered states
(\ref{eq-twoqubit-state-init-index-i}).
This property was theoretically justified in
\cite{amoi6-104,amoi6-118}
and confirmed by numerical tests in
\cite{amoi6-104,amoi6-118,amoi-arxiv-2021-sipqip}.
Its relevance may be outlined as follows.
For each expectation
$E \{
\Ytwoqubitsprobaindexstd
\}$
of a random probability
\ytwoqubitsprobaindexstd\ 
to be estimated, 
in practice the expectation operator
$E \{ . \}$
is replaced by a sample mean, i.e. by a \emph{sum} (of values,
moreover normalized). Similarly, each probability
\ytwoqubitsprobaindexstd\ 
is replaced by a sample frequency, i.e. by a \emph{sum} (of 1 and
0, depending whether the considered event occurs or not for
each trial defined by a preparation of the initial quantum
states 
(\ref{eq-twoqubit-state-init-index-i})
and by
an associated measurement 
{of the considered spin
component, for each of the two spins;}
this summation is
here again followed
by a normalization, by the total number of trials).
$E \{
\Ytwoqubitsprobaindexstd
\}$
is therefore estimated by a 
(normalized)
``sum of sums'', 
which may then be reinterpreted as a single
global
sum, and what primarily matters is the total number of 
preparations of initial quantum
states 
(\ref{eq-twoqubit-state-init-index-i})
involved in that global sum,
whereas the number of preparations for each state value
(\ref{eq-twoqubit-state-init-index-i})
may be decreased, down to 1.

In our previous papers, we applied the above analysis to the
probabilities
\ytwoqubitsprobaplusplus , 
\ytwoqubitsprobaplusminus\
and
\ytwoqubitsprobaminusminus\
of the data model
(\ref{eq-statecoef-vs-twoqubitsprobaplusplus-polar})-%
(\ref{eq-statecoef-vs-twoqubitsprobaminusminus-polar-vs-paramqubitindexstdstateplusmodulus}),
in order to achieve BQSS
\ytextmodifartitihundredsixtyfourversiononestepthree{\cite{amoi6-104,amoi-arxiv-2021-sipqip}}%
, BQPT 
\ytextmodifartitihundredsixtyfourversiononestepthree{\cite{amoi6-104,amoi6-118}}
and BHPE
\ytextmodifartitihundredsixtyfourversiononestepthree{\cite{amoi-arxiv-2021-sipqip}}%
.
Here, we apply it to a new SIPQIP task, namely 
intrusion detection,
thus introducing
its single-preparation version.
We therefore consider the expectations of
(\ref{eq-mixmodelstateoutqubitindexonedirectionnonerv-proba-plus-with-intrusion})
and
(\ref{eq-mixmodelstateoutqubitindexonedirectionnonerv-proba-plus-no-intrusion}).
For Case 1,
Eq.
(\ref{eq-mixmodelstateoutqubitindexonedirectionnonerv-proba-plus-with-intrusion})
thus yields
\begin{eqnarray}
E
\{
\Yprob
(
\Ymixmodelstateoutqubitindexonedirectionnonerv
=
+
)
\}
&
=
&
E
\{
\Yparamqubitonestateplusmodulus ^2 
\}
+
\Ytwoqubitresultphaseevolsin ^2
(
E
\{
\Yparamqubittwostateplusmodulus ^2
\}
-
E
\{
\Yparamqubitonestateplusmodulus ^2 
\}
)
\nonumber
\\
&
&
-
2 E
\{
\Yparamqubitonestateplusmodulus \Yparamqubittwostateplusmodulus
\sqrt{ 1
-
\Yparamqubitonestateplusmodulus ^2 } \sqrt{ 1
-
\Yparamqubittwostateplusmodulus ^2 } 
\sin \Ytwoqubitresultphaseinit
\}
\nonumber
\\
&
&
\hspace{8mm}
\times
\sqrt{ 1 -
\Ytwoqubitresultphaseevolsin ^2 } \Ytwoqubitresultphaseevolsin
.
\label{eq-mixmodelstateoutqubitindexonedirectionnonerv-proba-plus-with-intrusion-mean}
\end{eqnarray}
This might be further simplified
when moreover assuming that the RVs
$
\Yparamqubitonestateplusmodulus
,
\Yparamqubittwostateplusmodulus
$
and
$
\Ytwoqubitresultphaseinit
$
are statistically independent, as in
our previous BQSS and related investigations
\ytextmodifartitihundredsixtyfourversiononestepthree{(see e.g.
\cite{amoi6-67}
about random quantum sources and their independence)}%
.
However, that additional assumption is not required for the task
considered here.

Similarly, for Case 0,
Eq.
(\ref{eq-mixmodelstateoutqubitindexonedirectionnonerv-proba-plus-no-intrusion})
yields
\begin{equation}
E
\{
\Yprob
(
\Ymixmodelstateoutqubitindexonedirectionnonerv
=
+
)
\}
=
E
\{
\Yparamqubitonestateplusmodulus ^2 
\}
.
\label{eq-mixmodelstateoutqubitindexonedirectionnonerv-proba-plus-no-intrusion-mean}
\end{equation}

The two variants of the method of Section
\ref{sec-intrusion-multiple-prep}
may then be transposed to the single-preparation framework considered here.
\ytextmodifartitihundredsixtyfourversiononesteptwo{We hereafter
transpose
only the first variant, because it is especially attractive: It}
yields a simple 
protocol while
requesting Receiver 1 to have only limited prior knowledge about the
emitted
data, namely the value of
$
E
\{
\Yparamqubitonestateplusmodulus ^2 
\}
$.
This method may therefore be stated to be blind
(so-called blind signal processing methods are in fact
not completely blind because they
set some, although possibly very limited, conditions
on the considered data)
or semi-blind for the sake of clarity.
It should be noted that, unlike its multiple-preparation
version of 
Section
\ref{sec-intrusion-multiple-prep},
this single-preparation method does not require Receiver 1 to know any
\emph{individual} state value prepared by Emitter 1,
which is very attractive.
Moreover,
it does not depend on the
state values prepared by Emitter 2
(again provided the ``specific values'' are avoided).

This single-preparation
method operates as follows.
Receiver 1 gets a set of final two-qubit
states
\ymixsyststatefinal ,
without any request on the number of copies per
state value, 
unlike in Section
\ref{sec-intrusion-multiple-prep}.
After
performing a single one-qubit measurement 
on Qubit 1
for each such
state, 
Receiver 1
derives an estimate of
$
E
\{
\Yprob
(
\Ymixmodelstateoutqubitindexonedirectionnonerv
=
+
)
\}
$,
as explained above.
Receiver 1 then compares
this estimate 
to the known value
$
E
\{
\Yparamqubitonestateplusmodulus ^2 
\}
$
and makes the following decision,
based on
(\ref{eq-mixmodelstateoutqubitindexonedirectionnonerv-proba-plus-with-intrusion-mean})
and
(\ref{eq-mixmodelstateoutqubitindexonedirectionnonerv-proba-plus-no-intrusion-mean}):
If
$
E
\{
\Yprob
(
\Ymixmodelstateoutqubitindexonedirectionnonerv
=
+
)
\}
$
is ``close enough''
(with the same comment as above)
to
$
E
\{
\Yparamqubitonestateplusmodulus ^2 
\}
$,
then
Receiver 1 decides that no intrusion is being
performed (Case 0); otherwise, he decides that
intrusion is occurring (Case 1).
This method again exploits the fact that,
``in general'' (in the same sense as above),
$
E
\{
\Yprob
(
\Ymixmodelstateoutqubitindexonedirectionnonerv
=
+
)
\}
$
does not take the same value
in Cases 0 and 1, as shown
by
(\ref{eq-mixmodelstateoutqubitindexonedirectionnonerv-proba-plus-with-intrusion-mean})
and
(\ref{eq-mixmodelstateoutqubitindexonedirectionnonerv-proba-plus-no-intrusion-mean}).
\section{Discussion and conclusion}
\label{sec-concl}
A transform applied to a 
(possibly multi-qubit)
quantum state is often referred to as
a ``quantum process'', by the scientific community focused on
quantum process tomography,
or a ``quantum channel'',
by the scientific community focused on
quantum communications.
So far
in this paper, we focused on a particular class of such
quantum processes/channels,
namely two-qubit processes
based on
cylindrical-symmetry
Heisenberg-type exchange coupling.
This 
class of processes
is relevant for describing
coupling between two close electron spins, and that was our motivation
for investigating such processes
in our previous papers,
focused on
the field of spintronics
and on associated data processing tasks, such as BQSS, BQPT and BHPE.

When moving to the detection intrusion task with quantum channels
in the present paper,
we still considered the above class of channels
so far,
in order to more easily develop 
first 
concepts for intrusion detection,
by taking advantage of the information about these channels
that was already available from our previous investigations.
However, 
it should be clear that,
from the point of view of the detection intrusion task,
the above class of channels 
is here
only regarded as a toy model
for first investigations:
We
do not claim that it will be relevant
when then studying
practical communication scenarios, depending
on the considered hardware implementation.

\ytextmodifartitihundredsixtyfourversiononesteptwo{In particular,
communications based on photons deserve the following comments.
First considering the classical framework,}
\ytextmodifartitihundredsixtyfourversiononestepone{%
\ytextmodifartitihundredsixtyfourversiononesteptwo{everyday
communications}
use electromagnetic waves propagating either in
free space or in a solid 
\ytextmodifartitihundredsixtyfourversiononesteptwo{medium,}
e.g. an optical fiber
\ytextmodifartitihundredsixtyfourversiononestepthree{\cite{amansourelsevier2017}}%
. Such media are
non-magnetic, and their electric properties are classically described by the
induction vector $\overrightarrow{D}$ (H. Lorentz), representing a local
mean of the microscopic electric vector \cite{LandauVol8}. $\overrightarrow{E%
}$ being the applied field, in 
vacuum, $\overrightarrow{D}=\varepsilon _{0}%
\overrightarrow{E}$ 
(SI units, $\varepsilon 
_{0}$:
vacuum 
permittivity), and
in a dielectric medium $\overrightarrow{D}=\varepsilon _{0}\overrightarrow{E}%
+\overrightarrow{P}.$ The polarization $\overrightarrow{P}$ may be seen as
the response to the excitation $\overrightarrow{E}$, ferroelectrics, with a
spontaneous polarization, being an exception. Most dielectric media are
\ytextmodifartitihundredsixtyfourversiononesteptwo{\emph{linear,}}
i.e. $\overrightarrow{P}$ increases linearly with the excitation
(description using a scalar or more generally a tensor not depending upon
the excitation). 
\ytextmodifartitihundredsixtyfourversiononesteptwo{Moreover, the}
appearance of the laser in 1960, i.e. of intense
coherent electromagnetic sources, allowed the development of 
\ytextmodifartitihundredsixtyfourversiononesteptwo{\emph{non-linear}}
optics. Turning now to the quantum behavior associated with these phenomena,
one should 
\ytextmodifartitihundredsixtyfourversiononesteptwo{again}
make 
\ytextmodifartitihundredsixtyfourversiononesteptwo{a
distinction
bewteen linear and non-linear setups.}
One first thinks of an
electromagnetic wave propagating in 
vacuum space
(or possibly in a linear
medium): 
Already
in 1930 Dirac \cite{Dirac1930} considered a weak
electromagnetic 
\ytextmodifartitihundredsixtyfourversiononesteptwo{beam}
and its associated photons; a device separates this
beam into two partial beams, which are then made to interfere. One could
then a priori think that two distinct photons possibly interfere. But, in
such conditions, according to the general principles of quantum mechanics
\textquotedblleft \textit{each photon then interferes only with itself.
Interference between two different photons can never occur}%
\textquotedblright .}
\ytextmodifartitihundredsixtyfourversiononesteptwo{With respect to
photon-based quantum communications
addressed in the present paper, this entails that, if only
considering communications through 
vacuum space
or a linear medium,
no entanglement is created in the transmission channel itself.
This should be contrasted with
the scenarios 
\ytextmodifartitihundredsixtyfourversiononestepthree{considered above,}
where
entanglement is created by the channel itself (here with exchange coupling),
whereas the original two-qubit state associated with Emitter 1 and
Emitter 2 is unentangled.}

\ytextmodifartitihundredsixtyfourversiononestepone{%
\ytextmodifartitihundredsixtyfourversiononesteptwo{The above}
manifestation of the superposition principle
\ytextmodifartitihundredsixtyfourversiononesteptwo{for photons}%
, or
the presence of (previously prepared) entangled states when more than one
photon are implied may also be found in dielectrics, but other quantum
phenomena may also be found in some optically 
\ytextmodifartitihundredsixtyfourversiononesteptwo{\emph{non-linear}}
dielectric
materials:\ 1) two intense laser beams at frequencies $\omega _{1}$ and $%
\omega _{2}$\ may allow fluorescence at $\omega _{1}+\omega _{2};$ there,
two photons with respective frequencies $\omega _{1}$ and $\omega _{2}$
generate a photon with frequency $\omega _{1}+\omega _{2}\ $\cite{Boyd2008}.
2) spontaneous emission may sometimes allow emission at the difference
frequency: 
The
material receives a laser beam with frequency $\omega _{p}$
(p: pump), and emits at both $\omega $ (with a material dependent value) and 
$\omega _{p}-\omega $ (so-called optical parametric fluorescence); here, a
photon with frequency $\omega _{p}$ generates two photons, with respective
frequencies $\omega $ and $\omega _{p}-\omega $ \cite{Boyd2008}. Quantum
communications 
take profit of the superposition principle,
\ytextmodifartitihundredsixtyfourversiononestepthree{e.g.}
when
involving two photons in an entangled state, and of the no-cloning theorem,
both specific to quantum mechanics.\ Future quantum communication networks
should make use of\ 
quantum teleportation - which allows 
transport of
information, presently using entangled photon qubits - and of quantum
repeaters interconnecting quantum nodes \cite{Krenn2016}. An 
\ytextmodifartitihundredsixtyfourversiononesteptwo{eavesdropper,}
trying to access the information circulating within such a network could
e.g. try and operate 
\ytextmodifartitihundredsixtyfourversiononesteptwo{over}
a repeater.}
\ytextmodifartitihundredsixtyfourversiononesteptwo{Besides,
one may imagine a scenario involving}
transmission through quantum
channels, by means of photons,
be they 
initially entangled or not,
mainly with 
free propagation (linear medium), 
but now 
also with a 
\ytextmodifartitihundredsixtyfourversiononesteptwo{non-linear}
medium inserted
(e.g. by an eavesdropper)
in part of the overall transmission
path forming
what we called the ``main channel''
between Emitter 1 and Receiver 1 above.
One might
then
investigate to which extent an ``intruder'',
composed of Emitter 2 and Receiver 2, would
thus
be able to interact with the main channel so as to
extract information from it (eavesdropping), and to which
extent Receiver 1 would be able to detect this intrusion.
The 
relevance and attractiveness of this scenario need to
be further
investigated.

The 
competition between 
eavesdropping and intrusion
detection,
mentioned for the above new scenario, already
clearly appeared in the intrusion
detection
methods proposed in this paper:
these methods can detect intrusion ``in general'', i.e. except for
specific values of the considered parameters (quantum states and
channel parameters). Another extension of this paper therefore consists
of analyzing these specific values in more detail, in order to
determine 
whether they allow the eavesdropper to defeat the intrusion detector,
while extracting useful information from the main channel.
More generally speaking, in this paper we focused on the
capabilities of the proposed approaches in terms of intrusion detection,
but the associated eavesdropping capabilities should also be
analyzed in
\ytextmodifartitihundredsixtyfourversiononestepthree{our}
future work.

Finally, if one e.g.
aims at developing intrusion detection 
methods that are statistical in the sense that
they are based on 
averages of measurement outcomes
(to estimate probabilities or their expectations),
the following feature,
specific to the most advanced
methods that we proposed in this paper, should be kept
in mind
because it is likely to remain of high interest in
future methods too.
Our general SIPQIP framework for quantum processing
makes it possible to use only one instance of each source state
(i.e. emitted
state),
and this
is especially attractive in communication
scenarios, because (i) the receiver of the main channel
(Receiver 1) should preferably not require any
control 
on the states provided by the emitter
of the main channel (Emitter 1) and (ii) anyway, that receiver
surely has no control on the states provided by the
intruder%
\ytextmodifartitihundredsixtyfourversiononestepthree{, i.e. jammer}
(Emitter 2).
This ability of our SIPQIP methods to operate with one 
instance of each state
is obtained
by using \emph{expectation} of probabilities,
the latter probabilities being random-valued 
because we consider
random 
\ytextmodifartitihundredsixtyfourversiononesteptwo{quantum pure}
states.
In contrast,
a drawback of
usual QIP methods
is that they
require many copies
of a single state (or many copies per
state, if considering several states)
to estimate the individual probabilities associated
with that state.

\section*{Funding}
Not applicable.

\section*{Conflict of interest}
The authors declare that they have no conflict of interest.

\bibliographystyle{spmpsci}      
\bibliography{biblio_yd}

\begin{thebibliography}{10}
\providecommand{\url}[1]{{#1}}
\providecommand{\urlprefix}{URL }
\expandafter\ifx\csname urlstyle\endcsname\relax
  \providecommand{\doi}[1]{DOI~\discretionary{}{}{}#1}\else
  \providecommand{\doi}{DOI~\discretionary{}{}{}\begingroup
  \urlstyle{rm}\Url}\fi

\bibitem{a593}
Abed-Meraim, K., Qiu, W., Hua, Y.: Blind system identification.
\newblock Proceedings of the IEEE \textbf{85}(8), 1310--1322 (1997)

\bibitem{amq-baldwin-physreva-2014}
Baldwin, C.H., Kalev, A., Deutsch, I.: Quantum process tomography of unitary
  and near-unitary maps.
\newblock Physical Review A \textbf{90}, 012110--1 to 012110--10 (2014)

\bibitem{amq75}
Blume-Kohout, R., Gamble, J.K., Nielsen, E., Mizrahi, J., Sterk, J.D., Maunz,
  P.: Robust, self-consistent, closed-form tomography of quantum logic gates on
  a trapped ion qubit.
\newblock arXiv:1310.4492v1  (16 Oct. 2013)

\bibitem{Boyd2008}
Boyd, R.: Non-linear Optics.
\newblock Academic Press, Cambridge, USA (2008)

\bibitem{amq45official}
Branderhorst, M.P.A., Nunn, J., Walmsley, I.A., Kosut, R.L.: Simplified quantum
  process tomography.
\newblock New Journal of Physics \textbf{11}, 115010+12 (2009)

\bibitem{amq30official}
Chuang, I.L., Nielsen, M.A.: Prescription for experimental determination of the
  dynamics of a quantum black box.
\newblock Journal of Modern Optics \textbf{44}(11-12), 2455--2467 (1997)

\bibitem{book-comon-jutten-ap}
Comon, P., Jutten, C.: Handbook of blind source separation. Independent
  component analysis and applications.
\newblock Academic Press, Oxford, UK (2010)

\bibitem{amoi6-67}
Deville, A., Deville, Y.: Concepts and criteria for blind quantum source
  separation and blind quantum process tomography.
\newblock Entropy \textbf{19}(7 (July 2017)), paper no. 311 (2017)

\bibitem{amoi6-48}
Deville, Y.: Wiley Encyclopedia of Electrical and Electronics Engineering,
  chap. Blind source separation and blind mixture identification methods, pp.
  1--33.
\newblock Wiley, J. Webster (ed.) (2016)

\bibitem{amoi5-31}
Deville, Y., Deville, A.: Blind separation of quantum states: estimating two
  qubits from an isotropic {Heisenberg} spin coupling model.
\newblock In: Proceedings of the 7th International Conference on Independent
  Component Analysis and Signal Separation (ICA 2007), ISSN 0302-9743,
  Springer-Verlag, vol. LNCS 4666. {\it Erratum: replace two terms $ E \{ r_i
  \} E \{ q_i \} $ in (33) of \cite{amoi5-31} by $ E \{ r_i q_i \} $, since $
  q_i $ depends on $ r_i $.}, pp. 706--713. London, UK (2007)

\bibitem{amoi6-18}
Deville, Y., Deville, A.: Classical-processing and quantum-processing signal
  separation methods for qubit uncoupling.
\newblock Quantum Information Processing \textbf{11}(6), 1311--1347 (2012)

\bibitem{amoi6-34}
Deville, Y., Deville, A.: A quantum-feedforward and classical-feedback
  separating structure adapted with monodirectional measurements; blind qubit
  uncoupling capability and links with {ICA}.
\newblock In: Proceedings of the 23rd IEEE International Workshop on Machine
  Learning for Signal Processing (MLSP 2013). Southampton, United Kingdom
  (2013)

\bibitem{amoi6-37}
Deville, Y., Deville, A.: Blind qubit state disentanglement with quantum
  processing: principle, criterion and algorithm using measurements along two
  directions.
\newblock In: Proceedings of the 2014 IEEE International Conference on
  Acoustics, Speech, and Signal Processing (ICASSP 2014), pp. 6262--6266.
  Florence, Italy (2014)

\bibitem{amoi6-42}
Deville, Y., Deville, A.: Blind Source Separation: Advances in Theory,
  Algorithms and Applications, chap. 1. Quantum-source independent component
  analysis and related statistical blind qubit uncoupling methods, pp. 3--37.
\newblock Springer, Berlin, Germany, G. R. Naik and W. Wang Eds (2014)

\bibitem{amoi6-46}
Deville, Y., Deville, A.: From blind quantum source separation to blind quantum
  process tomography.
\newblock In: Proceedings of the 12th International Conference on Latent
  Variable Analysis and Signal Separation (LVA/ICA 2015), pp. 184--192.
  Liberec, Czech Republic, Springer International Publishing Switzerland, LNCS
  9237 (2015)

\bibitem{amoi6-104}
Deville, Y., Deville, A.: Stochastic quantum information processing, with
  applications to blind quantum system identification and source separation.
\newblock In: Proceedings of the 2018 IEEE 28th International Workshop on
  Machine Learning for Signal Processing (MLSP 2018),. Aalborg, Denmark (2018)

\bibitem{amoi-arxiv-2021-sipqip}
Deville, Y., Deville, A.: Single-preparation unsupervised quantum machine
  learning: concepts and applications.
\newblock https://arxiv.org/abs/2101.01442  (2021)

\bibitem{amoi6-118}
Deville, Y., Deville, A.: Quantum process tomography with unknown
  single-preparation input states: Concepts and application to the qubit pair
  with internal exchange coupling.
\newblock Physical Review A \textbf{101}(4), 042332--1 to 042332--18 (April
  2020)

\bibitem{amoi6-64}
Deville, Y., Deville, A.: Blind quantum source separation: quantum-processing
  qubit uncoupling systems based on disentanglement.
\newblock Digital Signal Processing \textbf{67}, 30--51 (August 2017)

\bibitem{book-equalization-ding-li}
Ding, Z., Li, Y.: Blind equalization and identification.
\newblock Marcel Dekker, New York (2001)

\bibitem{Dirac1930}
Dirac, P.: The Principles of Quantum Mechanics.
\newblock Clarendon Press, Oxford, GB (1930)

\bibitem{icabook-oja}
Hyvarinen, A., Karhunen, J., Oja, E.: Independent Component Analysis.
\newblock Wiley, New York (2001)

\bibitem{Krenn2016}
Krenn, M., Malik, M., Scheidl, T., Ursin, R., Zeilinger, A.: Optics in our
  time, M. D. Al-Amri, M. M. El-Gomati, and M. S. Zubairy Eds, chap. Quantum
  communication with photons, pp. 485--482.
\newblock Springer, Cham, Switzerland (2016)

\bibitem{LandauVol8}
Landau, L., Lifchitz, E.: Electrodynamique des milieux continus (French
  Edition, Physique th\'{e}orique, Tome 8).
\newblock Editions Mir, Moscow, USSR (1969)

\bibitem{amansourieice2000}
Mansour, A., Barros, A.K., Ohnishi, N.: Blind separation of sources: Methods,
  assumptions and applications.
\newblock IEICE Transactions on Fundamentals of Electronics, Communications and
  Computer Sciences \textbf{E83-A}(8), 1498--1512 (August 2000)

\bibitem{amansourelsevier2017}
Mansour, A., Mesleh, R., Abaza, M.: New challenges in wireless {\&} free space
  optical communications.
\newblock Elsevier Optics and Lasers in Engineering \textbf{89}, 95--108
  (February 2017)

\bibitem{amq50-physical-review}
Merkel, S.T., Gambetta, J.M., Smolin, J.A., Poletto, S., C\'orcoles, A.D.,
  Johnson, B.R., Ryan, C.A., Steffen, M.: Self-consistent quantum process
  tomography.
\newblock Physical Review A \textbf{87}, 062119--1 to 062119--9 (2013)

\bibitem{amq59}
Navon, N., Akerman, N., Kotler, S., Glickman, Y., Ozeri, R.: Quantum process
  tomography of a {M}\o lmer-{S}\o rensen interaction.
\newblock Physical Review A \textbf{90}, 010103--1 to 010103--5 (2014)

\bibitem{booknielsen}
Nielsen, M.A., Chuang, I.L.: Quantum computation and quantum information.
\newblock Cambridge University Press, Cambridge, UK (2000)

\bibitem{preskill-http-chap3}
Preskill, J.: Lecture notes for ph219/cs219: Quantum information and
  computation, ch. 3.
\newblock
  http://www.theory.caltech.edu/\symbol{126}preskill/ph219/chap3\_15.pdf  (2018
  (fall term))

\bibitem{livreproakis}
Proakis, J.G.: Digital communications.
\newblock McGraw-Hill, Boston (2001)

\bibitem{book-scharf}
Scharf, L.L.: Statistical signal processing. Detection, estimation and time
  series analysis.
\newblock Addison-Wesley, Reading, Massachusetts (1991)

\bibitem{amq48}
Shukla, A., Mahesh, T.S.: Single-scan quantum process tomography.
\newblock Physical Review A \textbf{90}, 052301--1 to 052301--6 (2014)

\bibitem{amq52-physical-review}
Takahashi, M., Bartlett, S.D., Doherty, A.C.: Tomography of a spin qubit in a
  double quantum dot.
\newblock Physical Review A \textbf{88}, 022120--1 to 022120--9 (2013)

\bibitem{amq56}
Wang, Y., Dong, D., Petersen, I.R., Zhang, J.: An approximate algorithm for
  quantum hamiltonian identification with complexity analysis.
\newblock In: Proceedings of the 20th World Congress of the International
  Federation of Automatic Control (IFAC 2017), pp. 12241--12245. Toulouse,
  France (2017)

\bibitem{amq41}
White, A.G., Gilchrist, A.: Measuring two-qubit gates.
\newblock Journal of the Optical Society of America B \textbf{24}(2), 172--183
  (Feb. 2007)

\end{thebibliography}

\end{document}